\documentclass[prl,twocolumn,showpacs,preprintnumbers,amsmath,amssymb]{revtex4-1}

\usepackage{graphicx}% Include figure files
\usepackage{dcolumn}% Align table columns on decimal point
\usepackage{bm}% bold math
\usepackage{amsmath}
\usepackage{latexsym}
\usepackage{hyperref}
\usepackage{color}

\newcommand{\rfig}[1]{Fig.~\ref{#1}}
\newcommand{\rref}[1]{Ref.~\cite{#1}}
\newcommand{\req}[1]{Eq.~(\ref{#1})}

\begin{document}

\title{Absence of a transport signature of spin-orbit coupling in graphene with indium adatoms}

\author{Zhenzhao Jia, Baoming Yan, Jingjing Niu, Qi Han, Rui Zhu}
\author{Xiaosong Wu}
\email{xswu@pku.edu.cn}
\author{Dapeng Yu}
\affiliation{
State Key Laboratory for Artificial Microstructure and Mesoscopic Physics, Peking University, Beijing 100871, China \\
Collaborative Innovation Center of Quantum Matter, Beijing 100871, China }

%\date{\today}

\begin{abstract}
Enhancement of the spin-orbit coupling in graphene may lead to various topological phenomena and also find applications in spintronics. Adatom absorption has been proposed as an effective way to achieve the goal. In particular, great hope has been held for indium in strengthening the spin-orbit coupling and realizing the quantum spin Hall effect. To search for evidence of the spin-orbit coupling in graphene absorbed with indium adatoms, we carry out extensive transport measurements, \textit{i.e.}, weak localization magnetoresistance, quantum Hall effect and non-local spin Hall effect. No signature of the spin-orbit coupling is found. Possible explanations are discussed.
\end{abstract}

\pacs{72.25.Rb 72.80.Vp 73.43.-f 81.05.ue}

\maketitle

The intrinsic spin orbit coupling (SOC) in graphene is extremely weak\cite{Huertas-Hernando2006,Min2006a,Yao2007a}. Enhancement of the coupling may give rise to a variety of topological phenomena, such as the quantum spin Hall effect (two dimensional topological insulators)\cite{Kane2005b,Kane2005a,Weeks2011,Hu2012,Jiang2012,Shevtsov2012}, quantum anomalous Hall effect\cite{Qiao2010,Ding2011,Tse2011,Qiao2012,Zhang2012,Qiao2013} and Chern half metals\cite{Hu2014}. These phenomena are among the hottest topics in condensed matter physics. Moreover, graphene endowed with strong SOC can have potential use in spintronics, as SOC provides a means to control the spin electrically, which is at the heart of spintronics.

Absorption of adatoms has been theoretically proposed as an effective way to enhance SOC in graphene\cite{CastroNeto2009,Ding2009a,Abdelouahed2010,Qiao2010,Ding2011,Tse2011,Weeks2011,Dyrdal2012,Hu2012,Jiang2012,Ma2012a,Qiao2012,Shevtsov2012,Zhang2012,Qiao2013,Gmitra2013,Ferreira2014,Hu2014,Pachoud2014}. By distorting the carbon $sp^2$ bond\cite{CastroNeto2009,Abdelouahed2010}, breaking the inversion symmetry\cite{Abdelouahed2010,Qiao2010,Weeks2011}, or mediating the hopping between the second-nearest-neighbours\cite{Kane2005b,Weeks2011}, intrinsic or Rashba SOC can be enhanced or induced. The intrinsic SOC is required for the predicted quantum spin Hall effect, whereas Rashba SOC destroys it\cite{Kane2005b}. It has been proposed that if the outer shell electrons of adatoms derive from $p$ orbitals, the induced intrinsic SOC always dominates over the induced Rashba interaction. Under this condition, it is possible to realize two dimensional(2D) topological insulators in graphene\cite{Weeks2011}. The most promising candidates are indium and thallium, which can open up a significant topologically nontrivial gap. Further theoretical work has confirmed that the two systems are indeed stable topological insulators\cite{Jiang2012,Shevtsov2012}.

Two experimental groups have reported angle-resolved photoemission studies on the spin-orbit splitting in a related system, graphene on metal substrates\cite{Varykhalov2008,Dedkov2008,Rader2009,Marchenko2012}. Graphene on gold displays a very strong Rashba effect. On the other hand, it has been found that the spin relaxation rate measured by non-local spin valves is not enhanced by gold adatoms\cite{Pi2010}, suggesting SOC is negligible. Recently, a strong SOC has been observed in hydrogenated graphene and chemical vapor deposited graphene by the spin Hall effect (SHE)\cite{Balakrishnan2013,Balakrishnan2014}. Nevertheless, in sharp contrast to numerous theoretical work on this topic, relevant experimental results, especially transport experiments, are scarce. This is in part due to two issues. One is related to the low diffusion barrier for metal adatoms\cite{Chan2008}, which causes clustering of adatoms at room temperature. The other is oxidation of adatoms.

In this work, we employ an ultra low temperature magnetotransport measurement system, with \textit{in situ} thermal deposition capability, to circumvent the two aforementioned issues. We choose indium, as it is reckoned by a few theoretical work as an ideal candidate\cite{Weeks2011,Jiang2012,Shevtsov2012}. Weak localization (WL), quantum Hall effect (QHE) and non-local SHE measurements have been carried out for different indium coverages with the aim of searching for evidence of SOC. Comparison with relevant theories has been made and no signature of SOC has been found. The implications have been discussed.

Graphene flakes were exfoliated from Kish graphite onto 285 nm SiO$_2$/Si substrates. Standard e-beam lithography and metallization processes were used to make Hall bar structures. Electrodes are made of 5 nm Pd/ 80 nm Au. Samples were annealed in Ar/H$_2$ atmosphere at 260 $^\circ$C for 2 hours to remove photoresist and other chemical residues and then transferred into our dilution refrigerator. The system is a modified Oxford dilution refrigerator, in which \textit{in situ} thermal deposition can be performed\cite{Parker2006}. Before the first deposition, current annealing was done to clean the surface. During deposition and measurements, the sample temperature was maintained below 5 K. Thus, the thermal diffusion of adatoms was strongly suppressed. Electrical measurements were done by a standard low frequency lock-in technique.

\begin{figure}[htb]
\includegraphics[width=0.48\textwidth]{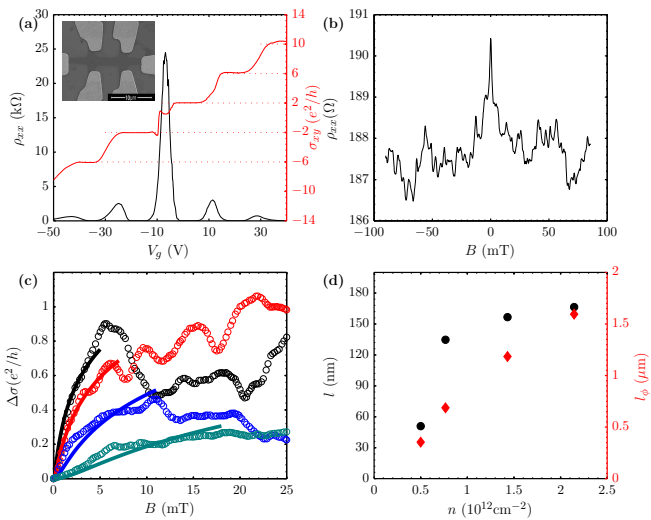}
\caption{\label{fig:basic} Magnetotransport of a graphene Hall bar device before indium deposition. (a) Longitudinal resistivity and the transverse conductivity versus the gate voltage in 9 T at 150 mK. The inset is a scanning electron microscopy picture of the device. (b) Low field magnetoresistivity exhibits two features, the weak localization peak at $B=0$ and the universal conductance fluctuations. (c) Fits to \req{eq:Mc2012} for the low field magnetoresistivity at different carrier densities, $n=2.15,1.43,0.77,0.50 \times 10^{12}$ cm$^{-2}$. (d) The mean free path $l$ and the phase coherence length $l_\phi$ as a function of $n_s$.}
\end{figure}

The sample geometry can be seen in the scanning electron microscopy image in \rfig{fig:basic}(a). The half integer QHE is well developed, which confirms that the sample is monolayer graphene. The low field magnetoresistivity is plotted in \rfig{fig:basic}(b). The narrow negative magnetoresistivity peak at $B=0$ is WL, while the noise-like but reproducible fluctuations are universal conductance fluctuations. In graphene, electrons are chiral and have a Berry phase $\pi$, which inverts the constructive interference to a destructive one. Thus, intrinsic graphene should display weak anti-localization (WAL). But, in the presence of intervalley scattering, resulting from short range potential, there will be a crossover from suppressed WL to WAL as the field increases\cite{McCann2006,Wu2007,Tikhonenko2008}. When the intervalley scattering rate exceeds the phase coherence rate, the low field WL correction to the conductivity can be expressed as\cite{McCann2012}:
\begin{equation}
\begin{split}
\Delta\sigma=-\frac{e^2}{4\pi ^2\hbar}\left[F\left(\frac{B}{B_\phi}\right)-F\left(\frac{B}{B_{\phi}+2B_\text{asy}}\right)\right.\\
\left. -2F\left(\frac{B}{B_{\phi}+B_\text{asy}+B_\text{sym}}\right)\right], \\
F(z)=\ln z + \psi\left(\frac{1}{2}+\frac{1}{z}\right), B_{\phi,\text{asy},\text{sym}}=\frac{\hbar}{4De\tau_{\phi,\text{asy},\text{sym}}}.
\label{eq:Mc2012}
\end{split}
\end{equation}
where $\psi$ is the digamma function, $e$ the elementary charge and $\hbar$ the reduced Planck constant. $D$ is the diffusion constant and $\tau_\phi$ is the phase coherence time. $\tau_\text{asy}$, $\tau_\text{sym}$ are the $z\rightarrow -z$ asymmetric and symmetric spin-orbit scattering time, respectively. For pristine graphene, SOC is negligible. To establish the baseline for later comparison, we have measured WL at different gate voltages (carrier densities), shown in \rfig{fig:basic}(c). Data are fitted to \req{eq:Mc2012} with only one parameter $\tau_\phi$. To meet the low field requirement of \req{eq:Mc2012} and also avoid the influence of the universal conductance fluctuations, only the low field positive magnetoconductance are fitted. A good agreement with the theory is found. The mean free path $l$ is calculated from the resistivity and carrier density. Considering the charge puddles in graphene, the carrier density at the Dirac point is taken as $0.5\times 10^{12}$ cm$^{-2}$\cite{Adam2007}. $l$ and the phase coherence length $l_\phi$ are plotted in \rfig{fig:basic}(d). $l_\phi$ decreases as one approaches the Dirac point. This is because $\tau_\phi$ in graphene is determined by electron-electron interaction, which is enhanced when screening is weakened. The suppression of $\tau_\phi$ is further enhanced by the reduction of the mean free time $\tau$\cite{Abrahams1981}.

\begin{figure}[htb]
\includegraphics[width=0.46\textwidth]{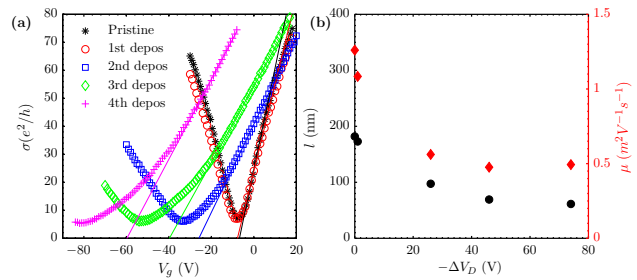}
\caption{\label{fig:depos} Deposition of indium. (a) The conductivity $\sigma $ versus gate voltage $V_g$ curves for the device after each deposition. The solid lines are linear fits, from which the field effect mobility is obtained. (b) The dependence of the mean free path $l$ and mobility $\mu$ on the shift of the Dirac point $\Delta V_\text{D}$.}
\end{figure}

Indium deposition was performed \textit{in situ} at a very slow rate for several times, each lasting 22 - 300 seconds. During deposition, the sample resistance was monitored so that a desired shift of the Dirac point could be obtained. After each deposition, electrical measurements were carried out. The conductivity $\sigma$ as a function of gate voltage is plotted in \rfig{fig:depos}(a). The Dirac point gradually shifts to negative gate voltage as the indium coverage increases, indicating electron doping. At the same time, the conductivity turns from sublinear to linear in $V_\text{g}$. The linear dependence is attributed to charged impurity scattering being dominant\cite{Adam2007}. So, the transition to the linear dependence suggests that indium adatoms mainly introduce charged impurities (long range potential), rather than short range potential. The minimum conductivity $\sigma_\text{min}$ at the Dirac point remains relatively constant, about 6$e^2/h$, while a closer look shows a slight decrease with increasing adatom density. A similar dependence of $\sigma_\text{min}$ has also been observed in potassium absorbed graphene\cite{Chen2008}. According to a self-consistent theory proposed by Adams \textit{et al.}\cite{Adam2007}, $\sigma_\text{min}$ is a consequence of two competing effects of charged impurities. One is to scatter electrons. The other is to generate a residue carrier density at the Dirac point by doping. The result is a weak negative dependence of $\sigma_\text{min}$ on the impurity density. At the same, the width of the $\sigma_\text{min}$ plateau increases, which is  observed in our experiment. So, all features in the density dependence of the conductivity are consistent with charged impurity scattering. Its implication on SOC will be discussed later.

We now estimate the area density of indium adatoms $n_\text{In}$. Assume that each indium adatom transfers $Z$ electrons to graphene. If adatoms are dilute, $Z$ should be a constant\cite{Adam2007}. Then, the doped carrier density $\overline{n}=Zn_\text{In}$. $\overline{n}$ can be estimated from the shift of the Dirac point $\Delta V_\text{D}$, as $\overline{n}=c_\text{g} \Delta V_\text{D}/e$. Here $c_\text{g}$ is the gate capacitance for 285 nm SiO$_2$ dielectric. The only uncertainty is the value of $Z$. According to first-principles calculations, $Z$ for indium on graphene is $0.8 \sim 1$\cite{Ribeiro2005,Chan2008,Weeks2011}. To get an idea of the coverage, we adopt $Z = 1$ to obtain its lower bound. Consequently, the area density after the third deposition is $3.1 \times 10^{12}$ cm$^{-2}$, corresponding to a coverage of 0.25\%.

From the gate dependence of the conductivity, the field effect mobility $\mu$ is obtained. Its dependence on $\Delta V_\text{D}$, which is proportional to $n_\text{In}$, is plotted in \rfig{fig:depos}(b), as well the mean free path $l$ at a carrier density of $3.8 \times 10^{12}$ cm$^{-2}$. As the mobility is substantially reduced after deposition, it is evident that adatom scattering dominates. We now look for signature of SOC induced by adatoms. WAL has been employed as a sensitive probe for SOC\cite{Hikami1980,Bergmann1982}. In conventional 2D electron gases with absence of SOC, the magnetoconductance is positive, the Hallmark of WL, stemming from constructive interference of electrons along time reversal paths. When SOC is turned on, it rotates the electron spin and produces destructive interference, giving rise to WAL, a negative magnetoconductance. W(A)L can be seen as a time-of-flight experiment. Specifically, interactions of a longer time scale manifest themselves in a lower magnetic field\cite{Bergmann1984}. So, as SOC increases, WAL first emerges from zero field and eventually dictates the whole field regime. The conductance correction is given by the HLN equation\cite{Hikami1980}:
\begin{equation}
\begin{split}
\sigma(B)=-g_sg_v\frac{e^2}{2\pi^2\hbar}\left[\psi\left(\frac{1}{2}+\frac{B_1}{B}\right)-\psi\left(\frac{1}{2}+\frac{B_2}{B}\right)\right.\\
\left.+\frac{1}{2}\psi\left(\frac{1}{2}+\frac{B_3}{B}\right)-\frac{1}{2}\psi\left(\frac{1}{2}+\frac{B_4}{B}\right)\right].
\end{split}
\label{eq:HLN}
\end{equation}
where
\begin{align*}
B_1&=B_0+B_\text{so}+B_\text{s}\\ B_2&=\frac{4}{3}B_\text{so}+\frac{2}{3}B_\text{s}+B_\phi\\
B_3&=2B_\text{s}+B_{\phi}\\
B_4&=\frac{2}{3}B_\text{s}+\frac{4}{3}B_\text{so}+B_{\phi}
\end{align*}
Here $B_0=\hbar/4De\tau$, $B_{\text{so},\text{s},\phi}=\hbar/4De\tau_{\text{so},\text{s},\phi}$. $\tau_\text{so}$, $\tau_\text{s}$, represent spin-orbit scattering time and magnetic scattering time, respectively. Since there are no magnetic impurities in our system, we neglect magnetic scattering.

In graphene, the expected evolution of magnetoconductance with increasing SOC is qualitatively similar. The reason is that, although intrinsic graphene display WAL, opposite to conventional 2D electron gases, typical graphene films show WL due to presence of defects. From the theory in \rref{McCann2012}, the magnetoresistance is given by \req{eq:Mc2012}\cite{McCann2012}. Compared with conventional 2D electron gases, the effect of SOC on WL depends on symmetry. For $z \rightarrow -z$ asymmetric SOC, normal crossover from WL to WAL occurs, while for $z \rightarrow -z$ symmetric SOC, WL will be suppressed. For adatom absorbed graphene, if any induced SOC, the $z \rightarrow -z$ asymmetric component should be substantial\cite{McCann2012}. It is anticipated that the magnetoconductance goes from negative to positive as the magnetic field increases. Therefore, both conventional 2D electron gases and graphene are predicted to show similar non-monotonic magnetoconductance. This is the feature that we are particularly interested in.

\begin{figure}[htb]
\includegraphics[width=0.46\textwidth]{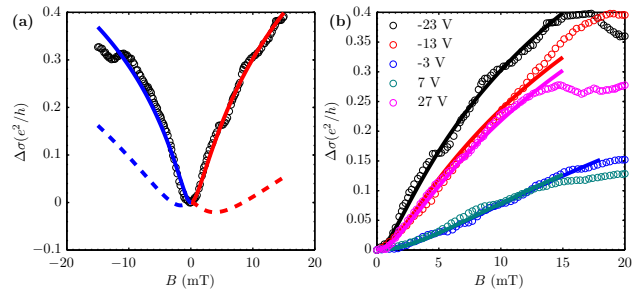}
\caption{\label{fig:WAL} Low field magnetoconductivity after the third deposition. (a) Fits of the low field magnetoconductivity to \req{eq:Mc2012} and \req{eq:HLN}. The circles are experimental data. The solid lines are the best fits to the equations, red for \req{eq:Mc2012} and blue for \req{eq:HLN}. The dotted lines are the plot of two equations, assuming a spin-orbit scattering time $\tau_\text{so}=\tau_\phi$. (b) Magnetoconductivity data and fits to \req{eq:Mc2012} at different gate voltages relative to the Dirac point ($V_g-V_\text{D}$).}
\end{figure}

\rfig{fig:WAL}(a) shows the low field magnetoconductance after the third deposition. The magnetoconductance monotonically increases with field, except for universal conductance fluctuations. No trace of WAL near $B=0$ has been found. Fitting of the data to \req{eq:Mc2012} yields $\tau_\phi=8.6$ ps, while $\tau_{asy}$ and $\tau_\text{sym}$ are an order of magnitude larger than $\tau_\phi$ with significant standard deviations, which essentially suggests inappreciable SOC. We have also performed fitting to \req{eq:HLN}. The obtained $\tau_\phi$ is similar, $\approx 10.9$ ps. Again, $\tau_\text{so}$ is much larger than $\tau_\phi$, consistent with \req{eq:Mc2012}. To illustrate the expected influence of SOC, both equations are plotted with $\tau_\phi$ obtained by fitting and all spin-orbit scattering time being equal to $\tau_\phi$. The resultant non-monotonic magnetoconductance is distinct from the experiment data. In fact, extensive measurements of the magnetoconductance at various carrier densities and after each deposition have been carried out and none of them shows WAL around $B=0$ (See supplementary materials). As a comparison, we have also performed the same experiments on deposition of magnesium, which is too light to induce appreciable SOC(See supplementary materials). Qualitatively similar results have been obtained, which confirms absence of induced SOC by In.

Since $\tau_\phi$ can be seen as a cut-off time for the quantum interference, it is reasonable to estimate that $\tau_\text{so}$ is longer than $\tau_\phi$ at least. For Elliott-Yafet spin-orbit scattering, which is most likely the case for adatoms,  $\tau_\text{so}=(E_\text{F}/\Delta_\text{so})^2\tau$. The upper-bound of the spin-orbit coupling strength $\Delta_\text{so}$ is then estimated as 12 meV at a carrier density of $1.67\times10^{12}$ cm$^{-2}$. We emphasize that this is the local SOC strength at an adatom, but not the overall spin-orbit gap of 7 meV at a 6\% coverage calculated in \rref{Weeks2011}. Because the gap approximately linearly diminishes with the coverage, the upper bound obtained in our experiment is actually much smaller than the prediction.

\begin{figure}[htb]
\includegraphics[width=0.46\textwidth]{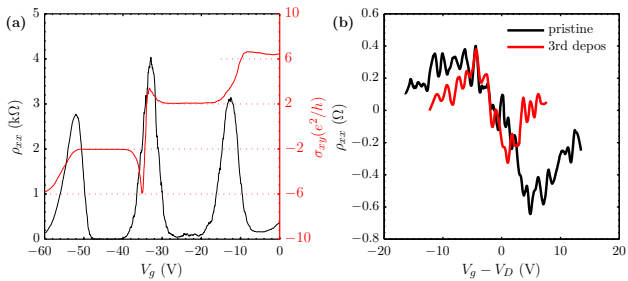}
\caption{\label{fig:others} Quantum Hall effect and SHE measurements. (a)Quantum Hall effect at $B=14$ T after the second deposition. (b) Non-local SHE.}
\end{figure}

Having not been able to observe SOC through WL, we turn to the quantum Hall effect. The famous half integer Hall effect in graphene stems from the Berry phase of $\pi$. In the presence of Rashba SOC, the electron spin is locked to its momentum\cite{Rashba2009,Zhai2014}. This spin texture adds another $\pi$ to the Berry phase\cite{Zhai2014}, as in topological insulators. The additional phase is expected to modify the quantum Hall effect. A theory proposes that the spin degeneracy in all Landau levels will be lifted\cite{Rashba2009}. Since there are two zero modes, one from the $n=0$ level and the other from $n=1$, the first quantum Hall plateaus stay at $\pm 2e^2/h$ (a spacing of $4e^2/h$), while the rest are spaced by $2e^2/h$. We have measured the quantum Hall effect after the second deposition, shown in \rfig{fig:others}(a). Despite a reduced mobility, the quantum Hall effect is evident and $n=0, \pm1$ quantum Hall plateaus remain at their original positions. The predicted lift of the degeneracy for $n=\pm1$ levels is not observed.

Another effect that may arise because of SOC is SHE. In a spin-orbit coupled system, a charge current generates a spin transport in the transverse direction, called SHE, and vice versa, called reverse SHE. The cooperation of two effects leads to a non-local resistance\cite{Abanin2011a},$
R_\text{nl}=\frac{1}{2}\left(\frac{\beta_\text{s}}{\sigma}\right)^2\frac{W}{\sigma l_\text{s}}e^{-L/l_\text{s}}$, where $\beta_\text{s}$ is the spin Hall conductivity, $l_\text{s}$ is the spin diffusion length, $L$ and $W$ are the length and width of the sample, respectively. This effect can be used to detect SOC. A large SOC in hydrogenated graphene and chemical vapor deposited graphene has been experimentally confirmed by this method\cite{Balakrishnan2013,Balakrishnan2014}. Here, we have measured the non-local resistance by injecting current through one pair of Hall probes of the Hall bar while monitoring the voltage signal across the other pair of Hall probes. The non-local resistance as a function of the gate voltage before and after deposition is plotted in \rfig{fig:others}(b). There is 0.6 Ohm non-local resistance before deposition. This resistance is caused by Ohmic contribution which decays as $e^{-\pi L/W}$. After deposition, no substantial change has been observed, indication of no appreciable induced SOC. Taking 0.6 Ohm as the upper-bound of the non-local resistance due to SHE and $\beta_\text{s}/\sigma \approx 0.45$ at a carrier density of $1\times 10^{12}$ cm$^2$/Vs from \rref{Balakrishnan2013} for hydrogenated graphene, we estimate $\tau_\text{so}=37$ ps, \textit{i.e.} $\Delta_\text{so}=$ 1.3 meV. This is an order of magnitude smaller than the 12 meV upper-bound estimated by WL. It should be pointed out that the estimation here is crude in that $\beta_\text{s}/\sigma$ is apparently a function of the SOC and unlikely the same as hydrogenated graphene.

Whereas the theories have listed indium as an important candidate for enhancing SOC in graphene and realizing a 2D Topological insulator, we fail to find any signature of SOC by transport measurements. It is noteworthy that the potential of adatoms has been theoretically treated as a short range one, as SOC is induced by mediating the hopping between the first, second and third nearest neighbours\cite{Pachoud2014}. However, the carrier density dependence of the conductivity doesn't support considerable increase of short range scattering. Similar observations have been made for magnesium (see supplementary materials) and potassium\cite{Chen2008}. The absence of induced SOC may be associated with the lack of short range scattering. The reason for the potential being short range can be accounted for by the Coulomb potential of ionized adatoms, which is long range. We notice that a recent study have shown that titanium particles dope graphene and give rise to long range scattering. But, when these particles are oxidized, accompanied by diminishing doping, significant short range scattering appears\cite{McCreary2011}. This implies that long range potential could ``screen" short range potential. In the presence of a strong long range potential, electrons will have less chance to get close enough to experience the local SOC, which will reduce its average strength. Another possibility is that the bond between indium adatoms and graphene is Van der Waals in nature. The interaction is too weak to modify the hopping between neighbours. Further study may focus on elements that induce less charge transfer, such as Fe or can form a stronger bond to graphene.

\begin{acknowledgments}
This work was supported by National Key Basic Research Program of China (No. 2012CB933404, 2013CBA01603) and NSFC (project No. 11074007, 11222436, 11234001). X. W. thanks P. Xiong for providing details of his \textit{in situ} deposition design.
\end{acknowledgments}

\end{document}